\documentstyle[12pt, aasms4]{article}
\righthead{CI in Evolved Stars}
\begin{document}
\title{Atomic Carbon in the Envelopes of Carbon-Rich Post-AGB Stars}

\author{G.R. Knapp}
\affil{Department of Astrophysical Sciences, Princeton University,
Princeton,  NJ 08544; gk@astro.princeton.edu}

\author{M. Crosas and K. Young}
\affil{Harvard-Smithsonian Center for Astrophysics, 60 Garden St.,
Cambridge, MA 02138;  rtm@dolson.harvard.edu; mcrosas@cfa.harvard.edu}

\and

\author{\v{Z}eljko Ivezi\'{c}}
\affil{Department of Astrophysical Sciences, Princeton University,
Princeton,  NJ 08544; ivezic@astro.princeton.edu}

\begin{abstract}

Atomic carbon has been detected in the envelopes of three carbon-rich
evolved stars: HD 44179 (=AFGL 915, the `Red Rectangle'); HD 56126;
and, tentatively, the carbon star V Hya.  This brings to seven the 
number of evolved star envelopes in which CI has been detected.  
Upper limits were found for several other stars,  including R CrB. 
CI was not detected in several oxygen rich post asymptotic giant branch (AGB) 
stars (OH231.8+4.2, for example), although it is detected
in their carbon-rich analogues.  Two trends are evident in the data.
First, circumstellar envelopes with
detectable CI are overwhelmingly carbon rich, suggesting that much
of the CI is produced by the dissociation of molecules other than CO.
Second, the more evolved the envelope away from the AGB, the higher is the 
CI/CO ratio.  The oxygen-rich supergiant star $\alpha$ Ori remains 
the only oxygen rich star with a wind containing detectable CI.

These data suggest an evolutionary sequence for the CI/CO ratio in
cool circumstellar envelopes.  This ratio is small (a few \%) while
the star is on the AGB, and the CI is located in the outer envelope and
produced by photodissociation.  The
ratio increases to about 0.5 as the star evolves away from the AGB
because of the dissociation of CO and other carbon-bearing molecules
by shocks caused by the fast winds which appear at the end of 
evolution on the AGB.  Finally, the ratio becomes $>>$ 1 as the
central star becomes hot enough to photodissociate CO.

\end{abstract}

\section{Introduction}

This paper reports a search for the 609 $\rm \mu m$ (492.1607 GHz)
$\rm ^3P_1 ~ \rightarrow ~ ^3P_o$ line of CI in the envelopes 
of ten evolved stars, with detections in three (HD 44179, HD 56126,
and, tentatively, V Hya).  These observations provide insight 
into the evolutionary status of the stars.

The CI(1--0) line is a useful probe of the
cold interstellar medium (Phillips and Huggins 1981, Keene 1995)
because it originates in almost all cases in photodissociation
regions in which the dominant gas-phase carrier of carbon changes from
CO to CII via CI.  It is widely observed from the photodissociation
regions in dense molecular clouds which are adjacent to star formation
regions and from the diffuse edges of molecular clouds where the
interstellar radiation field dissociates the molecular species in the
cloud (Keene 1995).

CI(1--0) emission has also been observed in several circumstellar
envelopes around luminous evolved late-type stars.  These envelopes are 
produced by mass loss from cool red giant and supergiant stars, and 
chemical equilibrium calculations show that gas-phase carbon is
probably entirely associated into CO and other molecules
in the stellar atmosphere, 
and hence in the material leaving the stars.
Circumstellar envelopes are a
particularly interesting type of molecular cloud because the radial
distribution in the envelope contains information both on the mass
loss history of the star and on the effects of photochemistry due
to the diffuse interstellar radiation field (Glassgold 1996).
Further, the central stars are evolving rapidly, and much of the
circumstellar material is destined to be ionized as the star evolves
away from the Asymptotic Giant Branch (AGB) and towards the white dwarf
stage. 

CI(1--0) emission has to date been detected in three circumstellar
envelopes  and four planetary nebulae,
and its likely origin is different in different cases:

(1) CI(1--0) emission has been detected from planetary nebulae which still
contain part of their circumstellar material in molecular form: AFGL 618
(Young 1997) whose CI content is roughly equal to that of CO; NGC 7027
(Young et al. 1999), also with a similar ratio of CI/CO: and the Ring and Helix
nebulae (Bachiller et al. 1994; Young et al. 1997), where n(CI) $>>$ n(CO).
The CI in these objects is likely to be mostly due to photodestruction
of CO by hot ultraviolet photons from the central star (Young 1997).
AFGL 618 is a very young planetary nebula; its central star is of type
B0 and most of its mass is in neutral form.  NGC 7027 is more evolved and
has a much hotter central star, but more than 50\% of its mass is 
still in neutral form, while the Ring and Helix nebulae have very hot
central stars and almost all of their mass in ionized form (Huggins et al. 
1996).  All of these nebulae have C/O $>$ 1 (Loup et al. 1993 and references
therein; Cox et al. 1998; Kholtygin 1998; Kwitter and Henry 1998).

(2) The CI(1--0) emission from the oxygen-rich supergiant $\alpha$ Ori
(Betelgeuse) shows that the CI abundance is about five times as high as
the CO abundance (Huggins et al. 1994; van der Veen et al. 1998). The hydrogen
in this envelope is also mostly or entirely in atomic form (Bowers and Knapp
1987) due to the star's relatively high effective temperature
(3500 K) and partially-ionized chromosphere
(Newell and Hjellming 1982; Wirsich
1988).

(3) The CI emission from the nearby carbon star IRC+10216 (Keene et al. 1993;
van der Veen et al. 1998) appears to be distributed in shells of about $\rm
15''$ and $\rm 45''$ radius.  Keene et al. (1993)
show that it is likely that the
CI in this envelope is due to photodissociation of carbon-containing 
molecules ($\rm C_2H_2$ for the inner shell and CO for the outer 
shell) by the interstellar radiation field, and that the gas leaving the 
star is almost entirely molecular.

(4) CI has been detected in the  carbon-rich
Egg Nebula, AFGL 2688 (Young 1997).  The 
central star is hotter than AGB stars and
is evolving away from the AGB, but is not yet hot
enough to ionize the surrounding circumstellar material.

The envelopes in which CI emission has been detected are, with the exception
of IRC+10216, around stars hotter than about 2500 K - 3000 K, the typical
temperature of AGB stars.  They are also, though not entirely, carbon-rich.
Guided by these results, we made a search for CI emission at the Caltech
Submillimeter Observatory in ten circumstellar envelopes, with emphasis
on stars with temperatures higher than those on the AGB,
including some post-AGB stars.  We also included two evolved stars, R CrB
and $\alpha$ Her, which are known to be losing mass or to have lost
mass in the past, but from which CO emission has never been detected.
In addition, the CO(4--3) line was observed in most of these stars.

The observations and results are described in the next section. 
Section 3 uses models of the circumstellar envelopes to derive the 
CI content, and  the results for the individual stars are discussed in
detail.

The results of the present paper are combined with those from the
literature in Section 4 to discuss the evolution of AGB stars as
probed by the CI line. The conclusions are also given in this section.
Appendix A summarizes CI observations of evolved stars 
accumulated in 1994 - 1996.  CI emission was not detected in any
of these stars.

\section{Observations and Results}

\subsection{Observations}

The basic data for the 13 evolved mass losing stars for which the
CO(4--3) and/or CI(1--0) lines were observed
are listed in Table 1.  The positions 
were taken from the Hipparcos catalogue (Perryman 1997;
Perryman et al. 1997): from radio interferometry observations: or from the
SAO catalogue.  All of the positions have an accuracy better that $\rm 1''$.
Also listed is the spectral type of the star, usually taken from SIMBAD
or from Loup et al. (1993).  Column 5 contains the chemistry of the
circumstellar envelope (C = carbon star, O = `oxygen' star, with n(O)
$>$ n(C)) from Loup
et al. (1993), and Column 6 the distance in parsecs.  

The observations of the 492.1607 GHz ground-state fine structure ($\rm
^3P_1 ~-~ ^3P_o$) line of CI were made on the nights of March 24 - 28 1998
using the 10.4 meter Robert B. Leighton telescope of the Caltech 
Submillimeter Observatory on Mauna Kea, Hawai`i.  The weather was superb
throughout the observing run, with zenith opacities at 220 GHz of $\rm
\tau_o ~ \leq ~ 0.03$, corresponding to $\rm \tau_o(492 ~ GHz)
\leq ~ 1$.
The observations were made using a liquid helium cooled SIS junction 
receiver with a double-sided system temperature of about 200 K.  The
CI line was observed in the lower sideband with the image sideband
at +3 GHz.  The CI line lies in the wings of a strong atmospheric
water vapor line at 487 GHz, and in this configuration the
sidebands see more or less the same atmospheric opacity.  The 
correction for the different atmospheric opacities in the two sidebands 
was estimated to be about 3\% for excellent weather conditions, and
is small enough to be ignored.

The spectral lines were measured using an acousto-optic spectrograph 
(AOS) with a bandwidth of 500 MHz over 1024 channels.  The spectrometer
frequency and spectral resolution were calibrated using an internally-
generated frequency comb, and the velocity scale is corrected to the
Local Standard of Rest (LSR).  The spectral resolution was measured to be 
$\rm \sim 0.65 ~ km~s^{-1}$.

The telescope half-power beamwidth is $\rm 15''$ at 492 GHz.  The observations
of each star were made by chopping between the star position and an adjacent 
sky position with the secondary mirror, using a chop throw of $\rm 60''$ in
azimuth at a rate of $\sim$ 1 Hz.  Pairs of chopped observations were made with
the source placed alternately in each beam.  The spectral baselines resulting
from this procedure are linear to within the r.m.s. noise for all 
observations.  The temperature scale and atmospheric opacity were measured 
by comparison with a hot (room temperature) load.  The line temperatures were
corrected for the atmospheric opacity and the main beam efficiency
(measured to be 45\% at 461 GHz using continuum observations of Venus) to the 
Rayleigh-Jeans equivalent main-beam brightness temperature $\rm T_{MB}$.

Before the CI line was observed for each star, the receiver was tuned to
the frequency of the nearby CO(4--3)  line at 461.0408 GHz, to measure
the telescope pointing offsets using the CO emission either from the star
itself or from a nearby CO-bright star.  In the process, 
a CO(4--3) line
profile was obtained for each star.

\subsection{Results}

Not all of the 13 stars were observed in both lines.  Only the CO(4--3)
line profile of IRC+10216 was observed for comparison with the CI(1--0)
observations by Keene et al. (1993).  The CO(4--3) emission from RY Dra
is weak, and no attempt was made to observe
the CI(1--0) line.  The CO(4--3) observation of IRAS 20000+3239 was made
in worsening weather conditions, which ended the observing run before any
useful CI(1--0) data could be obtained.  CO(4--3) emission was not observed
towards R CrB or $\alpha$ Her; many observations of the
CO lines  towards these stars have been made (e.g. Wannier et al.
1990; Nyman et al. 1992),
always with negative results.  The other stars were observed in both lines.

After correction for the main beam efficiency of 45\%, the data for each star
were co-added and a linear baseline subtracted from regions of the spectrum
assumed to be free of line emission.  The results are given in Figures 1
and 2 and in Table 2.  Figure 1 shows the CO(4--3) and CI(1--0) line 
profiles for the three stars in which the C1(1--0) line
was detected, and Figure 2
the CO(4--3) line profiles for the stars in which CI emission was not
detected.  The velocity coverage of the CO(4--3) line profile 
in Figure 1 is insufficient
to show the 200 $\rm km~s^{-1}$ outflow from V Hya (cf. Knapp et al. 1997). 
Observations of the CI line from
V Hya made on different nights are consistent; both 
detected emission near the stellar velocity.  The line profile in
Figure 1 is the average of the data obtained on the two nights.
However, the detection of CI(1-0) emission from this star is
tentative; as Figure 1 shows, there is a second feature at -110
$\rm km~s^{-1}$ which suggests the presence of baseline irregularities
or of weak emission in an unidentified line.

Table 2 contains for each observation the r.m.s. noise in the 500 MHz AOS
observation, the integrated line intensity in $\rm K \times km~s^{-1}$,
and the peak line temperature, central velocity $\rm V_c$ and terminal
wind outflow velocity $\rm V_e$.  These last three quantities were found by 
fitting a parabolic line profile to the data.  No attempt was made to
fit the complex CO(4--3) emission line from V Hya.  The fit for the 
equally complex CO(4-3) line from OH231.8+4.2 was made to the inner part
of the line profile only.  The upper limits to the CI intensity for
undetected stars were found
by integrating the data over the velocity range of the CO emission.  Where
no CO emission is seen, the data were summed over $\rm \pm 15 ~
km~s^{-1}$ centered on the star's radial velocity.  For stars with broad
wings like OH231.8+4.2, the integration was carried out only 
over the velocity range of the 
bright inner part of the line profile.

The errors in the line intensity and peak temperature quoted in Table 2
are statistical errors only, i.e. they give the signal to noise ratio of the 
observations.  We assume systematic errors of 10\% in the calibration
of the antenna temperature.

The CO and CI integrated line intensities are compared in Figures 3 and 4.
The data plotted include those
from Table 2 and for several other evolved stars
for which both CO(4--3) and CI(1--0) observations have been made.
Data for NGC 7027, and for the Helix and Ring nebulae, are not 
included.  For
IRC+10216, the CO(4--3) data are from Table 2 and the CI(1--0) data from
Keene et al. (1993).  For AFGL 2688, the CO(4--3) data are from
Young et al. (1992) and the CI(1--0) data from Young (1997).
For o Cet, the CO(4--3) data are from Young (1995) and the CI(1--0)
data from van der Veen et al. (1998), adjusted for the relative beam
areas of the JCMT and CSO telescopes.  There are no published CO(4--3) 
observations of AFGL 618 and $\alpha$ Ori that we are aware of. 
The expected CO(4--3) flux as observed at the CSO for these stars
is estimated from models
which reproduce the intensity of the lower-lying CO lines: the CO(3--2)
and CO(2--1) lines for AFGL 618 from Gammie et al. (1989) and the
CO(2--1) line for $\alpha$ Ori observed by Huggins et al. (1994).
The resulting sample contains 
data for seven carbon stars (with six CI detections), and six oxygen stars
(with one CI detection).  

Figure 3 shows the CI(1--0) line intensity versus the CO(4--3)
intensity.  The error bars are the combination of the statistical
uncertainties (Table 2) and an estimated 10\% systematic 
uncertainty for both the CI and CO
data. The uncertainties were set at 30\% for the stars whose CO(4--3)
line intensities were estimated from lower-lying CO transitions.
Data for the stars in which CI is not detected are the
`measured' intensities (Table 2) with 1$\sigma$ statistical error bars.

The dotted line in Figure 3 passes through the origin and the data for 
IRC+10216, to illustrate a possible proportionality between CO and CI 
intensity when CI is produced by photodissociation.  Compared to the
data for IRC+10216, none of the CI non-detections is significant at
the 3$\sigma$ level, i.e. were there photodissociation-produced 
CI in these stars, it would not be detectable.  The other six stars in
which CI emission is seen lie well above the line; and it is unlikely that
the CI in these circumstellar envelopes is due to `external' 
photodissociation by the interstellar radiation field. 

Figure 4 shows the ratio of the CI(1--0) and CO(4--3) lines as a function of
the spectral and chemical type of the central star.  The
3$\sigma$ upper limits on the CI(1--0) line flux, divided by the 
CO(4--3) line flux, are shown for stars in which CO emission is not
detected.  Figure 4 suggests that the earlier the spectral type
of the central star, and the more carbon rich the envelope, the greater
the relative abundance of CI.  However, this trend is far from universal,
and Figure 4 suggests rather that there are several different mechanisms
which can produce detectable CI in a circumstellar envelope.

\section{The CI/CO Ratio}

The CO(4--3) and CI(1--0) emission from the envelopes was modeled using
a line excitation/radiative transfer code for a uniformly expanding,
constant mass loss rate, spherical envelope (Morris 1980, Knapp and
Morris 1985, Crosas and Menten 1997).  The level populations are determined 
by collisions with neutral species and, in the case of CO, by radiative
excitation via the 4.6$\rm \mu m$ v = 0$\rightarrow$1 line.  The CO
extent of the envelope is assumed to be truncated by the photodissociation 
of CO (Mamon, Glassgold and Huggins 1988).  The line profile as it would be
observed by a given telescope is calculated by convolving the emergent 
line intensity across the envelope with a circular gaussian model of the 
telescope beam.  The model requires knowledge of the distance, the wind 
velocity (which is measured from the CO profile, Table 2) and the 4.6 $\rm 
\mu m$ flux, found from the observations tabulated by Gezari et al. (1993).
Atomic carbon has three fine-structure levels in the ground state, $\rm 
^3P_o$, $\rm ^3P_1$ and $\rm ^3P_2$, which lie 0 K, 23 K and 62 K respectively 
above the ground state.  The 492 GHz line corresponds to the $\rm ^3P_1
\rightarrow ^3P_o$ transition.

The envelope modeling proceeded as follows.  First, the CI(1--0)
line was modeled: the free parameters are the mass loss rate, the $\rm
CI/H_2$ abundance, and the radial extent of the CI in the envelope.
The CO(4--3) line was modeled with a single variable parameter, the 
$\rm CO/H_2$ abundance.

Models of the three stars in which CI emission was detected, HD 44179, 
HD 56126, and V Hya, are discussed below in detail.  The 
results are summarized in Table 3.

\subsection{HD 44179, AFGL 915, the `Red Rectangle'}

This unique and remarkable object is a carbon-rich red biconical nebula
discovered by Cohen et al. (1975) with a complex red emission spectrum
(e.g. Waelkens et al. 1992) containing PAH and $\rm CH^+$ emission
(Balm and Jura 1992).  The CO line flux measured by Jura, Balm and
Kahane (1995) is far weaker relative to the star's 60$\rm \mu m$ 
flux density than is typical for other carbon stars (Olofsson et al. 1993).
The CO line profile has two components centered at the same velocity, one with
full width at zero power of about 4 $\rm km~s^{-1}$ and the other with a
width of about 12 $\rm km~s^{-1}$.  Jura et al. (1995) suggest that 
the broad component can be associated with the mass loss outflow and the 
narrow component with a disk which entrains the bipolar flow.  The star
is a spectroscopic binary, with both stars embedded in a dusty torus or
disk (van Winckel, Waelkens and Waters 1995).  The extent of the bipolar
outflow is $\rm \pm 40''$ (Waelkens et al. 1996), giving a dynamical
age (assuming a distance of 330 pc  and an outflow velocity
of 6 $\rm km~s^{-1}$) of $\rm \sim 10^4$ years.  This suggests that the disk
which entrains the flow is long-lived, as discussed by Jura et al. (1995,
1997). Strong
evidence for this hypothesis comes from the recent finding that the
chemistry of the Red Rectangle is mixed: the broad
component is carbon rich, while the narrow component
is oxygen rich (Waelkens et al. 1992; Balm and Jura 1992; Reese
and Sitko 1996; Waters et al. 1998).  These authors suggest that the 
oxygen-rich narrow component
is long lived and was formed during a previous phase of 
binary-enhanced mass loss; the star subsequently evolved to a carbon
star, which is now producing the bipolar outflow (cf. the discussion of 
BM Gem by Kahane et al. 1998).

The spectral type of HD 44179 is given by SIMBAD as B8 - A0, but the presence
of a small HII region detected via its radio 
frequency continuum emission (Knapp et al. 1995; Jura et al. 1997)
shows that there is a hotter component in the system, with a temperature
at least that of a B3 star, 25,000 K.  Given the peculiar nature of the 
Red Rectangle, and in particular its small central HII region and low
CO/60 $\rm \mu m$ flux ratio, it is an obvious candidate for a CI
search, and as Figure 1 and Table 1 show, CI emission was indeed detected.

Is the CI(1--0) emission associated with the narrow
or broad component seen in CO?  Figure 5 shows the CO(4--3) and CI(1--0) line
profiles from Figure 1, plotted together to allow the line shapes
to be compared.  The CO(4--3) line profile observed at the CSO, like the
CO(1--0) and CO(2--1) line profiles observed by Jura et al. (1995), has
two velocity components, with half-widths at 0 $\rm km~s^{-1}$
of $\sim$4 and $\sim$8.5 $\rm
km~s^{-1}$.  The CI(1--0) line width is $\sim$ 4 $\rm km~s^{-1}$,
and the line profile shape comparison suggests that the CI emission is
partly associated with both components, but that most of it is associated
with the narrow component.

The model of the Red Rectangle was made starting with the CI(1--0) emission.
The distance is poorly known: we assume the Cohen et al. (1975)
distance of 330 pc, which is consistent with the Hipparcos observation
of $\pi$ = 2.62$\pm$2.37 mas (Perryman et al. 1997).  The best fit to the 
CI(1--0) emission was found with $\rm \mathaccent 95 M ~ = ~ 3 \times 10^{-6}
 ~ M_{\odot} ~ yr^{-1}$.  Two models, one with CI/$\rm H_2$  = $\rm 2.5
\times 10^{-5}$ and R(CI) = $\rm 10^{17}$ cm, and the second with 
CI/$\rm H_2$ = $\rm 10^{-4}$, R(CI) = $\rm 1.5 \times 10^{16}$ cm, fit
the data reasonably well (see Figure 6), but the latter fit is 
somewhat better.  The narrow component of the CO(4--3) line ($\rm
T_{MB}$ = 0.8 K, $\rm V_e$ = 4 $\rm km~s^{-1}$) was then fit by varying 
the CO abundance, giving $\rm CO/H_2 ~ = ~ 7.5 \times 10^{-6}$.  The
CI/CO ratio in this component is thus about 13.  Note that this calculation 
assumes a simple spherical outflow model for both broad and narrow components.

We also modeled the broad ($\rm V_e ~= ~ 8.5 ~ km~s^{-1}$) component.
Although the signal-to-noise ratio of the CI(1--0) line does not rule out the 
presence of CI in this component, all of the gas phase C is 
assumed to be in
CO.  The mass loss rate for this component is then $\rm \sim 10^{-7}
 ~ M_{\odot} ~ yr^{-1}$, assuming that it is carbon-rich and that 
$\rm CO/H_2 ~ = ~ 10^{-3}$.
Thus, N(CI) $>>$ N(CO) in the Red Rectangle, suggesting that the
weakness of the CO emission relative to the 60$\rm \mu m$
emission is due to a small CO
abundance.  To check this, the IRAS and published submillimeter
continuum data were used to calculate the dust content of the envelope.
Walmsley et al. (1991) measured the 1.3 mm 
flux density with the 15 m SEST, while van der Veen et al. (1994)
measured the 450$\rm \mu m$, 800$\rm \mu m$ and 1.1 mm flux densities with
the 15 m JCMT.  The spectral index of these observations is -2.9, 
showing that the dust emissivity index at these wavelengths is 0.9
(the small HII region contributes negligibly to the total flux density at
these wavelengths).  The simple model of 
Knapp, Sandell and Robson (1993)was used, with $\rm L_{\star} ~ = ~ 10^3 ~
L_{\odot}$, $\rm V_o ~ = ~ 7.5 ~ km~s^{-1}$ and graphite grains, to find
$\rm \mathaccent 95 M(grains) ~ = ~ 4 \times 10^{-8} ~ M_{\odot}
 ~ yr^{-1}$.  The total gas to dust ratio in the envelope is then
$\sim$75 by mass. Most of the circumstellar gas and dust in the system
appears to be in the 4 $\rm km~s^{-1}$ component, which is identified
with the oxygen-rich circumstellar disk.
This estimate is highly uncertain and model dependent, but
provides a plausible fit to the envelope properties: we conclude that
most of the  
gas-phase C is in CI, the gas to dust ratio is more or less normal,
and the shallow emissivity index shows the presence of large grains.
However, the fact that CI emission is detected from the narrow component,
and that a very large majority of circumstellar shells in which CI
emission is detected are carbon rich, argues that the disk component
may also be carbon rich, rather than oxygen rich.

\subsection{HD 56126}

This object is likely to be a post-AGB star; its spectral type is F5Iab
(SIMBAD listings) and it has a circumstellar envelope whose infrared 
colors suggest that mass loss has ceased within the last few hundred years
(Kwok et al. 1990). Bright CO line emission (Bujarrabal
et al. 1992; Knapp et al. 1998) is seen from the envelope,
and its mid-infrared emission shows that it
is axisymmetric (Dayal et al. 1998).  The envelope is carbon
rich (Bakker et al. 1997) and the star itself has a variability period of
about 12 days (L\`ebre et al. 1996; Bakker et al. 1997) and is possibly
an RV Tau variable.  The mid-infrared spectrum shows strong emission in the 
Brackett series, blueshifted by about 1000 $\rm km~s^{-1}$ (Kwok et al. 1990),
showing the presence of a fast wind from the central star.
The distance is not known; we use 2.4 kpc, estimated from the radial 
velocity, 73 $\rm km~s^{-1}$, assuming Galactic rotation.
The CO(4--3) line
intensity in Table 2 is about four times that of the CO(2--1) line
observed at the CSO (Knapp et al. 1998), showing that the envelope is 
unresolved on a scale of $\rm 20''$. The CO lines are parabolic,
with no sign of a fast wind - the CO(2--1) observations reported by
Knapp et al. (1998) find no fast-moving molecular gas to a limit of about
2\% of the brightness of the peak CO line emission.

The CI and CO lines have the same velocity and velocity width ($\rm V_e
 ~ = ~ 10 ~ km~s^{-1}$, Table 2).  The CI line was modeled with $\rm
\mathaccent 95 M ~ = ~ 9.5 \times 10^{-6}~ M_{\odot} ~ yr^{-1}$,
CI/$\rm H_2$ = $\rm 4 \times 10^{-4}$, and R(CI) = $\rm 5 \times 10^{17}$
cm.  This model is compared with the data in Figure 7.
The CO line intensity can be reproduced by an abundance $\rm CO/H_2$
= $\rm 10^{-3}$, typical of values found for carbon stars (Lambert et al. 
1986).  At this mass loss rate and CO abundance, the radius at which half of 
the CO is photodissociated by the interstellar ultraviolet field is about
$\rm 4 \times 10^{17}$ cm, using the calculations of Mamon, Glassgold and
Huggins (1988). The derived CI and CO abundances are thus similar, the 
outflow velocities measured in both lines are the same, and the model CI
radius is similar to that at which CO photodissociates.  These results suggest
that the CI in the HD56126 envelope is produced by photodissociation 
of the outflowing CO in the envelope, perhaps aided by the axisymmetric
structure of the envelope.  However, as Figure 3 shows, the CI/CO line strength
ratio is far larger for HD 56126 than is that for IRC+10216, for which it is
reasonable to assume that the observed CI is due to photodissociated carbon 
containing molecules.
HD 56126 is too cool (spectral type F5)
to produce sufficient UV photons to dissociate the 
CO, and there is no evidence for a hotter component (although this
envelope has not been searched for radio continuum emission).
A more likely
source of the CI is collisional dissociation of slow-moving
circumstellar CO by the fast 
wind (Kwok et al. 1990). 

\subsection{V Hya}

The carbon star V Hya has bright CO emission, weak CS emission and a very
unusual CO line shape, with two horns and a Voigt-like line profile, quite
different from the parabolic, steep sided line profile seen from almost
all other AGB envelopes.  In addition, the star has a fast molecular wind
with an outflow velocity of at least 200 $\rm km~s^{-1}$ (Knapp et al.
1997).

The distance to V Hya can be estimated at 380 pc assuming a standard 
absolute K magnitude, consistent with the upper limit to the Hipparcos
parallax, 0.16$\pm$1.29 mas, Perryman et al. (1997).  The velocity and
spatial structure of the CO emission from the V Hya envelope is complex
(Kahane et al. 1996; Knapp et al. 1997).  The CS(5--4) and (7--6)
lines have relatively simple shapes, and the line widths suggest an
outflow velocity of 15 $\rm km~s^{-1}$, consistent with the width of the
weak, tentatively detected CI line (Table 2, Figure 1).  The CI emission
can be fit with $\rm \mathaccent 95 M ~ = ~ 1.5 \times 10^{-6} ~ M_{\odot}
 ~yr^{-1}$, CI/$\rm H_2 ~ = ~ 3 \times 10^{-4}$, and a CI shell radius
of $\rm 5 \times 10^{16}$ cm.  The fitted profile is compared with the data in
Figure 8.  The inner parts of the CO(4--3) profile
can then be fit with $\rm CO/H_2 ~ = ~ 10^{-3}$, giving a CI/CO ratio
of 0.3.

V Hya is a cool carbon star, and there is no evidence of a hotter 
component in the system.  It is thus unlikely that the CI is produced by 
photodissociation.  Given the presence of the fast molecular wind
and shock-excited optical line emission (Lloyd
Evans 1991), a more likely mechanism for producing the CI is shock
dissociation of circumstellar CO. The signal-to-noise ratio of the CI
line profile is far too low to detect any fast-moving CI.
J type shocks can dissociate CO, but do
not ionize carbon: the possibility of 
shock dissociation can be tested by searching for the [OI]63$\mu$m and 
[CII]158$\mu$m lines, or by the presence and line ratios of the $\rm 
H_2$ ro-vibrational lines (Hollenbach and McKee 1989).

\subsection{Non-Detections}

No CI emission was detected from any of the other stars.  The limit on the 
CI/CO ratio is not sufficient for most of these stars (Y CVn, RY Dra, X Her,
89 Her) to rule out a CI abundance similar to that of the detected stars
(Figure 3). Y CVn was observed because of its unusual chemistry (it
is a J-type carbon star) and because, like HD 44179, it has a low
ratio of $\rm I(CO)/S(60\rm \mu m)$.
The cases of OH231.8+4.2 and AFGL 2343 are more interesting.
OH231.8+4.2 is, perhaps, an oxygen-rich analogue to V Hya: it has a cool
central star and no internal source of photoionization, a fast molecular 
outflow, and shock-excited optical line emission (Reipurth 1987).  Unlike
V Hya, OH231.8+4.2 has much cooler IRAS colors and is oxygen rich.
The lack of detection of CI emission from OH231.8+4.2 is then consistent with
the weak emission from V Hya, given the different envelope chemistries.
AFGL 2343 is an oxygen-rich post-AGB
G5 supergiant, with a very high mass loss rate. The star is less evolved
than its closest carbon star analogue, AFGL 2688, and does not yet have a
fast molecular wind.  

Two stars which are known to be losing mass but have never been detected in
the CO line are $\alpha$ Her and R CrB.  The latter star is the
prototype of a very
rare class; carbon-rich, hydrogen-poor, yellow giants (the 
spectral type of R CrB is GOIab) which at random intervals undergo
a steep decline in brightness by many magnitudes
due to the expulsion of dust (see Clayton 1996).  R CrB
is surrounded by a large dust shell $18'$ in diameter (Gillett et al.
1986) whose mass may be several $\rm M_{\odot}$. 
It is a semiregular variable with several periods (Rao and Lambert
1997, Feast et al. 1997, Feast 1997), has an inner dust shell
also detected by IRAS, and has recently begun another fading episode 
(Walker et al. 1986). ISO observations show a featureless spectrum.
Several attempts to detect
CO emission from this star have proved unsuccessful (see Loup et al.
1993), as has a search for HI (Clayton 1996).  Given the high effective 
temperature (6500 K, Rao and Lambert 1997) any gas shed by the star
is expected to be atomic rather than molecular.

Considerable observing time was expended on this star; as well as the time
spent in March 1998, CI observations were obtained in March 1996.  No
emission was detected in the summed observations at an r.m.s. level 
of 0.016 K (Table 2), setting a 3$\sigma$ upper limit on the CI column density 
of about 2 $\rm \times 10^{15}$ cm.  The failure to detect CI 
emission from this star could have many causes, but one of the more
plausible explanations lies in the hydrogen deficiency of the star - 
the gas densities in any circumstellar gas may not be high enough
to collisionally excite the transition.

\section{Discussion and Conclusions}

In this paper, we describe a search for atomic carbon in the circumstellar
envelopes of evolved stars using the 492 GHz $\rm ^3P_1-^3P_0$
fine structure line.  CI was detected in three envelopes, those around 
HD 44179 (the Red Rectangle), HD 56126 and (tentatively) V Hya.  Not
counting evolved planetary nebulae, this brings to seven (including the
transition object AFGL 618) the total number of evolved star envelopes
in which CI emission has been detected.  CI was not detected in sixteen
other evolved stars (seven of them discussed above and a further nine in
Appendix A). Of the detected stars, V Hya (tentatively detected)
is at the earliest stage of evolution: it is still on the AGB and 
does not have  hot stellar component, but is ejecting
a fast molecular wind as well as a slow wind, and the CI may be produced
by shocks where the fast and slow winds interact.  The CI/CO ratio
in this envelope is about 0.3.  HD 56126 has an F5 spectral type, and
is a post-AGB star.  The CI/CO ratio is about 0.4.  The hottest of these
three stars is HD 44179, whose spectral type is B0 - B3; 
CI/CO $\sim$ 13 for this object.  These three stars show that a
progressively larger fraction of the carbon in a circumstellar
envelope is in the form of CI as the central star becomes hotter
and more evolved beyond the AGB.

Tables 3 and 4 summarize the observational results to date;
Table 3 lists the evolved stars from which CI emission has been
detected, and Table 4 the stars in which it is not detected.  Table
3 includes the four planetary nebulae in which CI emission has
been detected.

Tables 3 and 4 show several trends: (1) CI is overwhelmingly detected
in carbon rich envelopes.  In carbon and oxygen rich envelopes of
otherwise comparable properties (e.g. OH 231.8+4.2 and V Hya;
AFGL 2688 and IRC+10420; HD 56126 and 89 Her), CI is detected
in the carbon-rich envelope and not in the oxygen-rich envelope.
This suggests that much of the CI may originate in the destruction
of molecules other than CO - $\rm C_2, ~ C_2H_2$, for example - which
are easier to destroy. (2) in most stars, whatever CI is produced by
external photodissociation, i.e. by dissociation by the interstellar
radiation field, is either too small in quantity or is produced at too
large radii (where the total gas density is low) to be detectable.
(3) the presence of a hot ($\geq$ 20,000 K) central star very 
quickly produces significant photodissociation - the  very young
planetary nebulae AFGL 618 and HD 44179, both of which have hot central
stars, have CI/CO $\geq$ 0.5.  (4) in carbon rich post AGB stars such as
AFGL 2688, V Hya, and HD 56126, detectable CI is present even before the 
central star becomes very hot. 
The first two stars have 
fast molecular winds, the third a fast ionized stellar wind. 
The indirect circumstantial evidence suggests shock production
of CI.  (5) $\alpha$
Ori is, to date, unique  - its surface temperature is high enough that
the wind is largely atomic, but atomic gas is not detected from other
supergiants (CE Tau, VY CMa) in the observed sample.  (6) The gaseous
component of the copious mass loss that must occur for R CrB remains
undetected.

These observations show that the CI in the envelopes of evolved stars
has a variety of origins, as it does in other regions of the 
interstellar medium.  Almost all evolved stars are cool enough that
their winds are dusty and molecular, and CI is produced in these 
envelopes, as in other molecular clouds, by the shock- and/or 
photo-destruction of circumstellar molecules.
The data also show the dependence of the CI content on evolution - the
CI/CO ratio rises as the star evolves away from the AGB.

\acknowledgements

Some of the observations described in Appendix A were made in collaboration
with Karl Menten: we thank him for allowing us to publish those
observations here.
We are very grateful to the director of the CSO, T. G. Phillips, for
granting the observing time for this project, and to the staff for much
help and advice with the observations.  We thank the referee, Jocelyn Keene,
and the editor, Steve Willner, for very helpful comments.
This research made use of the SIMBAD data
base, operated at CDS, Strasbourg, France. Astronomical research at the 
C.S.O. is supported by the National Science Foundation via grant AST96-15025.
Support for this work from Princeton University and from the N.S.F. via
grant AST96-18503 is gratefully acknowledged.

\appendix
\section{Searches for CI in Evolved Stars, 1994--1996}

This Appendix summarizes negative results for searches for CI emission
from evolved stars made in 1994-1996 at the CSO.  These observations
were made before the CSO was equipped with its chopping secondary
mirror, and so are not as sensitive as subsequent observations.  The
status of the CSO for these observations, and the observing methods,
are as described by Young (1997).

The observations are summarized in Table 5, which gives the object, its
1950 position, its circumstellar chemistry, the type of object,
and the 5$\sigma$ upper limit to the brightness temperature in the 
CI(1--0) line.

The nine objects in Table 5 were observed for a variety of reasons.  M 1-16
and IRAS 17243-1755 are planetary nebulae.  M 1-16 is a bipolar planetary
nebula, while IRAS 17243-1755 is a young planetary nebula like AFGL 618
with a small central HII region embedded in a molecular envelope - it
is, however, oxygen rich.  Like $\alpha$ Her, $\alpha$ Sco is a red
giant star which is known to be losing mass, but which has not been
detected in the CO lines.  VY CMa and IRC+10420 are both oxygen-rich 
supergiant stars.  The central star of IRC+10420 is an F supergiant,
so this is a post-AGB star; like AFGL 2343, it may be an oxygen-rich
analogue of AFGL 2688.  R Leo is a nearby bright Mira for which CO
observations give a fairly low mass loss rate (1 -- 2 $\rm \times 10^{-7}
M_{\odot} ~ yr^{-1}$).  R Vir is a Mira variable of early spectral type.
Young (1995) showed that the cooler Miras (spectral type M 6 and later)
have circumstellar shells which are readily detected in the CO lines,
while the warmer Miras (M5 and earlier) do not: again, by analogy
with $\alpha$ Ori, these stars may be ejecting an atomic stellar wind.  
Finally, the only carbon-rich object in Table 5, CIT 6, was observed
because it is an evolved star with a carbon-rich circumstellar 
envelope whose CO emission is second in brightness only to 
that of IRC+10216.
The failure to detect CI in this star, and the small amount of 
(photoproduced) CI in IRC+10216, shows that when n(C)$>$n(O), the
excess carbon is in various carbon molecules rather than atomic
form.

\clearpage
\begin{deluxetable}{crrrrrrrrrrrrrrr}
\footnotesize
\tablecaption{Observed Stars: Basic Parameters \label{tbl-1}}
\tablewidth{0pt}
\tablehead{
\colhead{Star} & \colhead{$\alpha$(1950)} & \colhead{$\delta$(1950)} & 
\colhead{Sp.T.} & \colhead{Chem.} & \colhead{D(pc)} & \colhead{Refs}
}
\startdata
HD 44179& 06 17 37.0& $-$10 36 52& B0-B3& C& 330& 1\nl
HD 56126& 07 13 25.3& $+$10 05 09& F5& C& 2400& 2\nl
OH231.8+4.2& 07 39 58.9& $-$14 35 44& M6& O& 1300& 3\nl
IRC+10216& 09 45 14.8& +13 30 41& C& C& 150& 4\nl
V Hya& 10 49 11.3& -20 59 05& N: C7,5& C& 380& 5\nl
Y CVn& 12 42 47.1& +45 42 48& C5,5& C& 220& H\nl
RY Dra& 12 54 28.1& +66 15 52& C4,5& C& 500& H\nl
R CrB& 15 46 30.7& +28 18 32& F8I& C& & \nl
X Her& 16 01 08.8& +47 22 36& M6& O& 140& H\nl
$\alpha$ Her& 17 22 22.3& +14 20 46& M5& O& 120& H\nl
89 Her& 17 53 24.0& +26 03 24& F2Iab& O& 1000& H\nl
AFGL 2343& 19 11 24.9& +00 02 19& G5& O& & \nl
IRAS20000+3239& 20 00 02.8& +32 39 07& G8Ia& C& \nl

\enddata
\tablenotetext{} {Spectral types from SIMBAD listings unless otherwise
noted}

\tablenotetext{} {Chemistry from Loup et al. (1993)}

\tablenotetext{} {H = distance from Hipparcos parallax (Perryman 1997; 
Perryman et al. 1997) or from other sources, consistent with Hipparcos
parallax limit}

\tablenotetext{}{Other distances: (1) Cohen et al. 1995. (2) calculated
assuming Galactic rotation. (3) Bowers and Morris 1984. (4) see discussion
by Jura (1994). (5) calculated assuming standard bolometric K magnitude of -8.2
(e.g. Jura 1994).}

\end{deluxetable}

\clearpage

\begin{deluxetable}{crrrrrrrrrrrrrrr}
\footnotesize
\tablecaption{Observational Results. \label{tbl-2}}
\tablewidth{0pt}
\tablehead{
\colhead{Star} & \colhead{Line} & \colhead{r.m.s.} & \colhead{I} &
\colhead{$T_{MB}$}& \colhead{$\rm V_c$}&
\colhead{$\rm V_e$} & \colhead{Vel. Range}\nl
& & \colhead{(K)} & \colhead{(K $\rm km~s^{-1}$)} & \colhead{(K)} &
\colhead{$\rm (km ~ s^{-1}$)} & \colhead{$\rm (km ~ s^{-1}$)} &
\colhead{$\rm (km ~ s^{-1}$)} 
}
\startdata
HD 44179& CO(4--3)& 0.052& 7.26$\pm$0.17& 0.89$\pm$0.04& +0.2$\pm$0.2&
5.6$\pm$0.2& \nl
& CI(1--0)& 0.058& 1.92$\pm$0.16& 0.27$\pm$0.04& +0.3$\pm$0.3& 5.1$\pm$0.7& \nl
& & & & & & & \nl
HD 56126& CO(4--3)& 0.095& 36.1$\pm$0.5& 2.59$\pm$0.05& +72.9$\pm$0.4&
10.3$\pm$0.4& \nl
& CI(1--0)& 0.053& 2.82$\pm$0.23& 0.17$\pm$0.02& +74.1$\pm$1.0& 12.7$\pm$1.4&
\nl
& & & & & & & \nl
OH231.8+4.2& CO(4--3)& 0.100& 163.5$\pm$1.1& 2.89$\pm$0.03& +35.4$\pm$0.3&
29.1$\pm$0.5& 0--70\nl
& CI(1--0)& 0.065& $-$0.3$\pm$1.1& -& -& -& \nl
& & & & & & & \nl
IRC+10216& CO(4--3)& 0.60& 433.3$\pm$3.0& 20.1$\pm$0.2& $-$25.4$\pm$0.2&
15.0$\pm$0.1 \nl
& & & & & & & \nl
V Hya& CO(4--3)& 0.10& 127.3$\pm$1.1& -& -& -& -100--100\nl
& & & 102.9$\pm$1.1& -& -& -& -33,0\nl
& CI(1--0)& 0.038& 2.45$\pm$0.17& 0.11$\pm$0.02& $-$16.4$\pm$1.3& 16.4$\pm$1.5
& \nl
& & & & & & & \nl
Y CVn& CO(4--3)& 0.088& 17.3$\pm$0.3& 1.45$\pm$0.05& +21.7$\pm$0.1& 
9.5$\pm$0.4& \nl
& CI(1--0)& 0.053& 0.0$\pm$0.2& -& -& -& \nl
& & & & & & & \nl
RY Dra& CO(4--3)& 0.106& 9.5$\pm$0.4& 0.61$\pm$0.05& -5.0$\pm$0.4&
9.0$\pm$0.5& \nl
& & & & & & & \nl
R CrB& CI(1--0)& 0.016& 0.00$\pm$0.13& -& -& -& -20,80\nl
& & & -0.23$\pm$0.05& -& -& -& -10,15\nl
& & & & & & & \nl
X Her& CO(4--3)& 0.090& 21.7$\pm$0.4& 2.54$\pm$0.06& -73.1$\pm$0.2&
7.5$\pm$0.2& \nl
& CI(1--0)& 0.04& 0.0$\pm$0.2& -& -& -&  \nl
& & & & & & & \nl
$\alpha$ Her& CI(1--0)& 0.056& -0.017$\pm$0.20& -& -& -& \nl
89 Her& CO(4--3)& 0.084& 3.6$\pm$0.3& 0.47$\pm$0.07& -7.61$\pm$0.49& 
5.7$\pm$1.1& \nl
& CI(1--0)& 0.05& -0.37$\pm$0.20& -& -& -& \nl
& & & & & & & \nl
AFGL 2343& CO(4--3)& 0.101& 94.1$\pm$0.7& 1.79$\pm$0.03& +99.7$\pm$0.3&
36.3$\pm$0.4& \nl
& CI(1--0)& 0.059& -0.62$\pm$0.65& -& -& -& \nl
& & & & & & & \nl
IRAS20000+3239& CO(4--3)& 0.11& 7.2$\pm$0.4& 0.32$\pm$0.03& +12.6$\pm$1.1&
18.7$\pm$2.3& \nl
\enddata

\end{deluxetable}

\clearpage
\begin{deluxetable}{crrrrrrrrrrrrrrr}
\footnotesize
\tablecaption{CI/CO for Evolved Stars with Detected CI \label{tbl-3}}
\tablewidth{0pt}
\tablehead{
\colhead{Star} & \colhead{Chem} & \colhead{N(CI)/C(CO)} & 
\colhead{$\rm T_{eff}$} & \colhead{Refs}
}
\startdata
IRC+10216& C& 0.02& 1250& Y97& \nl
AFGL 2688& C& 0.07& 6500& Y97& \nl
AFGL 618& C& 0.7& 30,000& Y97& \nl
NGC 7027& C& 0.5& 170,000& Y97& \nl
NGC 6720& C& 10& 110,000& Y97& \nl
NGC 7293& C& 6& 90,000& Y87& \nl
$\alpha$ Ori& O& 5& 3,500& H94,V98& \nl
V Hya& C& 0.3& 2650& This paper& \nl
HD 56126& C& 0.4& 6500& This paper& \nl
HD 44179& C& 13& 25,000& This paper& \nl
\enddata
\tablenotetext{} {H94: Huggins et al. 1994}

\tablenotetext{} {Y97: Young 1997}

\tablenotetext{} {V98: van der Veen et al. 1998}

\end{deluxetable}

\clearpage
\begin{deluxetable}{crrrrrrrrrrrrrrr}
\footnotesize
\tablecaption{CI in Evolved Stars: Non-Detections\label{tbl-4}}
\tablewidth{0pt}
\tablehead{
\colhead{Star} & \colhead{Chem} & \colhead{Sp. Type} &
\colhead{Refs}
}
\startdata
Y CVn & C& C5,5& 1,2& \nl
R CrB & C & GOIab& 1& \nl
X Her& O& M6& 1& \nl
$\alpha$ Her& O& M5 & 1& \nl
89 Her & O& F2Iab& 1& \nl
AFGL 2343 & O& G5& 1& \nl
OH231.8+4.2& O& M6& 1& \nl
o Cet& O& M5.5e& 2& \nl
NML Tau& O& M8e& 2& \nl
CE Tau& O& M2Iab& 2& \nl
VY CMa& O& M5Iab& 2,3& \nl
R Leo & O& M7e& 2,3& \nl
SW Vir& O& M7III& 2& \nl
V CrB& C& C6,2e& 2& \nl
TX Psc& C& C7,2& 2& \nl
CIT 6& C& C4,3& 3& \nl
$\alpha$ Sco& O& M1.5Ib& 3&\nl
M 1-16& O& PN& 3& \nl
IRAS17423-1755& O& PN& 3& \nl
R UMi& O& M7& 3& \nl
IRC+10420& O& F8Ia& 3& \nl
R Vir& O& M4.5III& 3& \nl
\enddata
\tablenotetext{} {1. This work: observations with CSO}

\tablenotetext{} {2. van der Veen et al. 1998: observations with JCMT}

\tablenotetext{} {3. This Work, Appendix A: observations with CSO}

\end{deluxetable}

\clearpage
\begin{deluxetable}{crrrrrrrrrrrrrrr}
\footnotesize
\tablecaption{ Negative Results for CSO CI 
Observations, 1994-1996 \label{tbl-5}}
\tablewidth{0pt}
\tablehead{
\colhead{Star} & \colhead{$\alpha$(1950)} & \colhead{$\delta$(1950)} & 
\colhead{Chem.} & \colhead{Obj. Type} & \colhead{$\rm T_{MB}$}
}
\startdata
M 1-16& 07 34 55.2& -09 32 00& O& PN& $<$0.4& \nl
IRAS17423-1755& 17 42 18.9& -17 55 36& O& PN& $<$ 0.3& \nl
VY CMa& 07 20 54.7& -25 40 13& O& SG& $<$ 0.2& \nl
R UMa& 10 41 07.9& +69 02 20& O& Mira& $<$ 0.2& \nl
IRC+10420& 19 24 26.8& +11 15 11& O& SG/PAGB& $<$0.2& \nl
R Leo& 09 44 52.2& +11 39 42& O& Mira& $<$0.2& \nl
R Vir& 12 35 57.7& +07 15 48& O& Mira& $<$0.9& \nl
CIT 6& 10 13 10.7& +30 49 17& C& Mira& $<$0.4& \nl
$\alpha$ Sco& 16 26 20.3& -26 19 22& O& RG& $<$0.4& \nl

\enddata

\tablenotetext{} {PN = planetary nebula, Mira = Mira variable, SG = 
supergiant, PAGB = post-AGB star, and RG = red giant}

\end{deluxetable}

\clearpage
\figcaption{CO(4--3) and CI(1--0) line profiles observed with the CSO
for three evolved stars in which CI emission is detected. \label{fig1}}

\figcaption{CO(4--3) line profiles of eight evolved stars in which
CI was not detected. \label{fig2}}

\figcaption{Integrated intensity, in $\rm K \times km~s^{-1}$, of the CI(1--0)
line versus that of the CO(4--3) line observed with the Caltech Submillimeter
Observatory.  The open symbols are observations of oxygen-rich stars, the 
filled symbols of carbon-rich stars.  The dotted line suggests the 
proportionality between these two lines for stars in which the CI is produced 
by photodissociation. \label{fig3}}

\figcaption{Ratio of the intensities of the CI(1--0) and CO(4--3) lines as a
function of spectral type.  Inverted triangles are 3$\sigma$ upper limits
for the CI line for undetected stars.  Open symbols: oxygen stars.  Filled
symbols: carbon stars. \label{fig4}}

\figcaption{Comparison of the CO(4--3) (heavy line) and CI(1--0) (light
line) line profiles for HD 44179, the ``Red Rectangle''.  The intensity scale 
for the CO(4--3) line has been reduced by a factor of four. \label{fig5}}

\figcaption{Two models (heavy line) for the CI(1--0) emission from
HD 44179, the `Red Rectangle', compared with the data (light line).
Upper panel: $\rm CI/H_2 = 2.5 \times 10^{-5}$, R(CI) = $10^{17}$ cm.
Lower panel: $\rm CI/H_2 = 10^{-4}$, R(CI)=1.5 $\times 10^{16} $cm.
\label{fig6}}

\figcaption{Comparison of model CI emission (heavy line) with data
(light line) for HD 56126.\label{fig7}}

\figcaption{Comparison of model CI emission (heavy line) with data
(light line) for V Hya.\label{fig8}}

\clearpage

\plotone{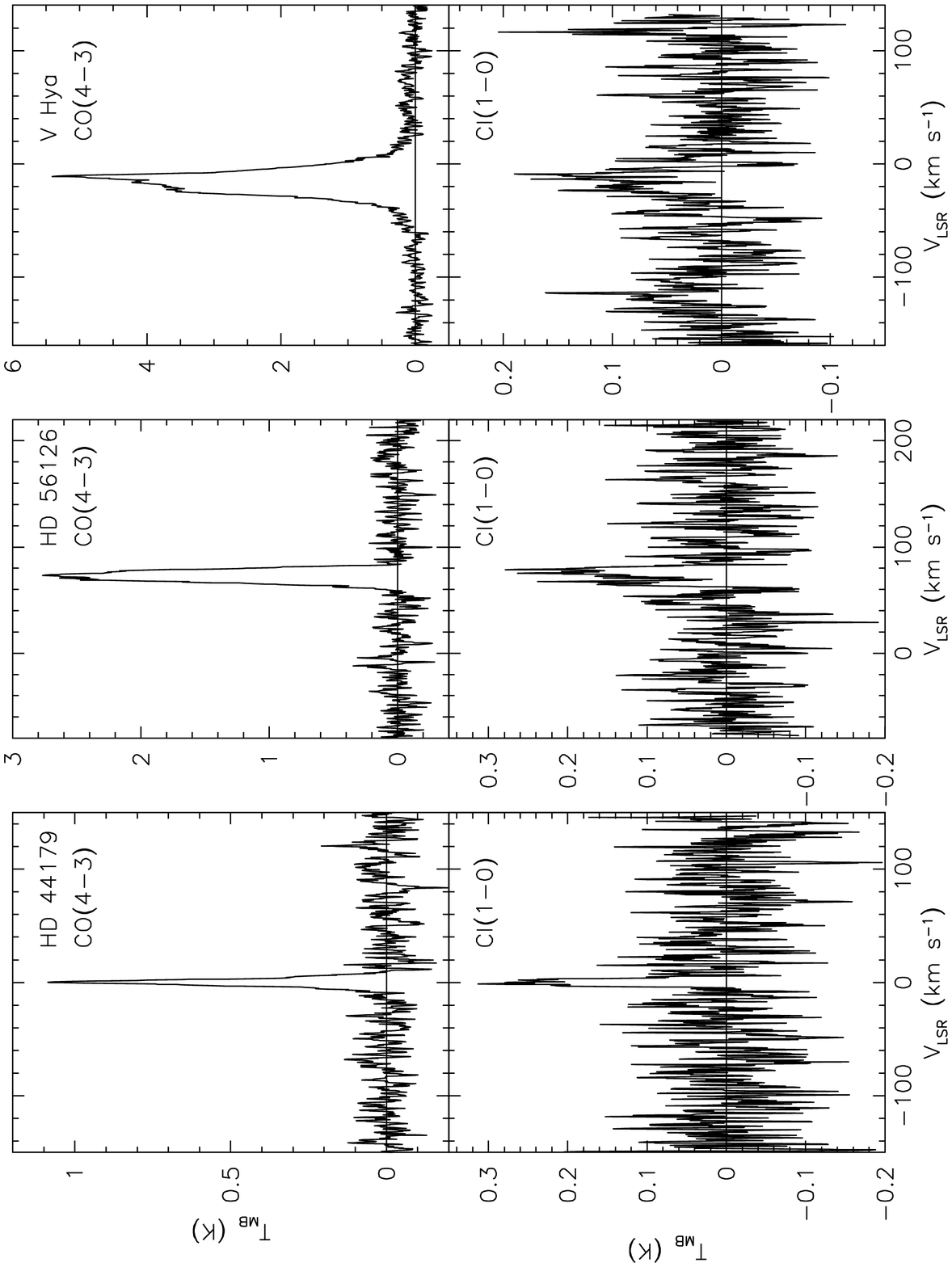}

\clearpage

\plotone{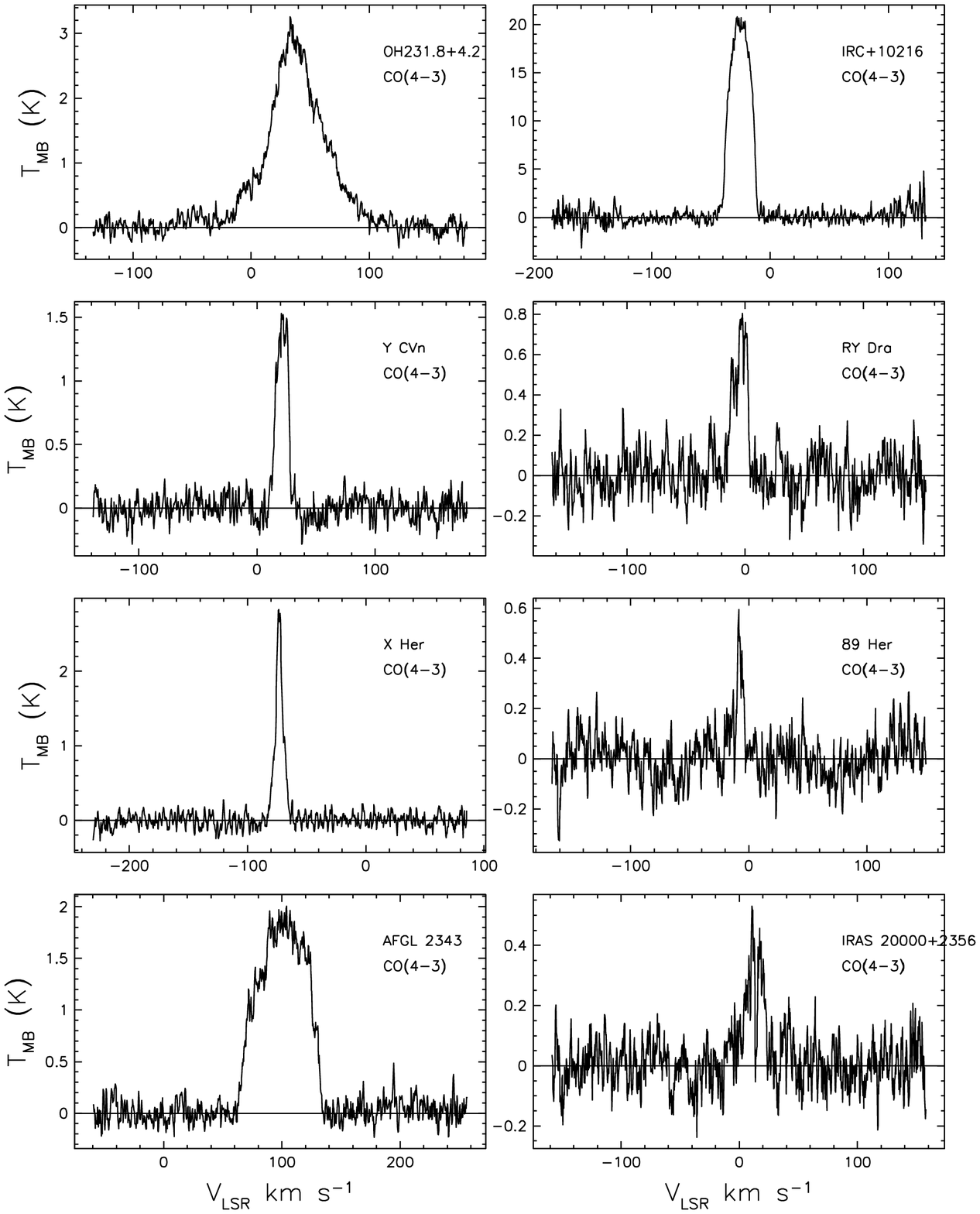}

\clearpage

\plotone{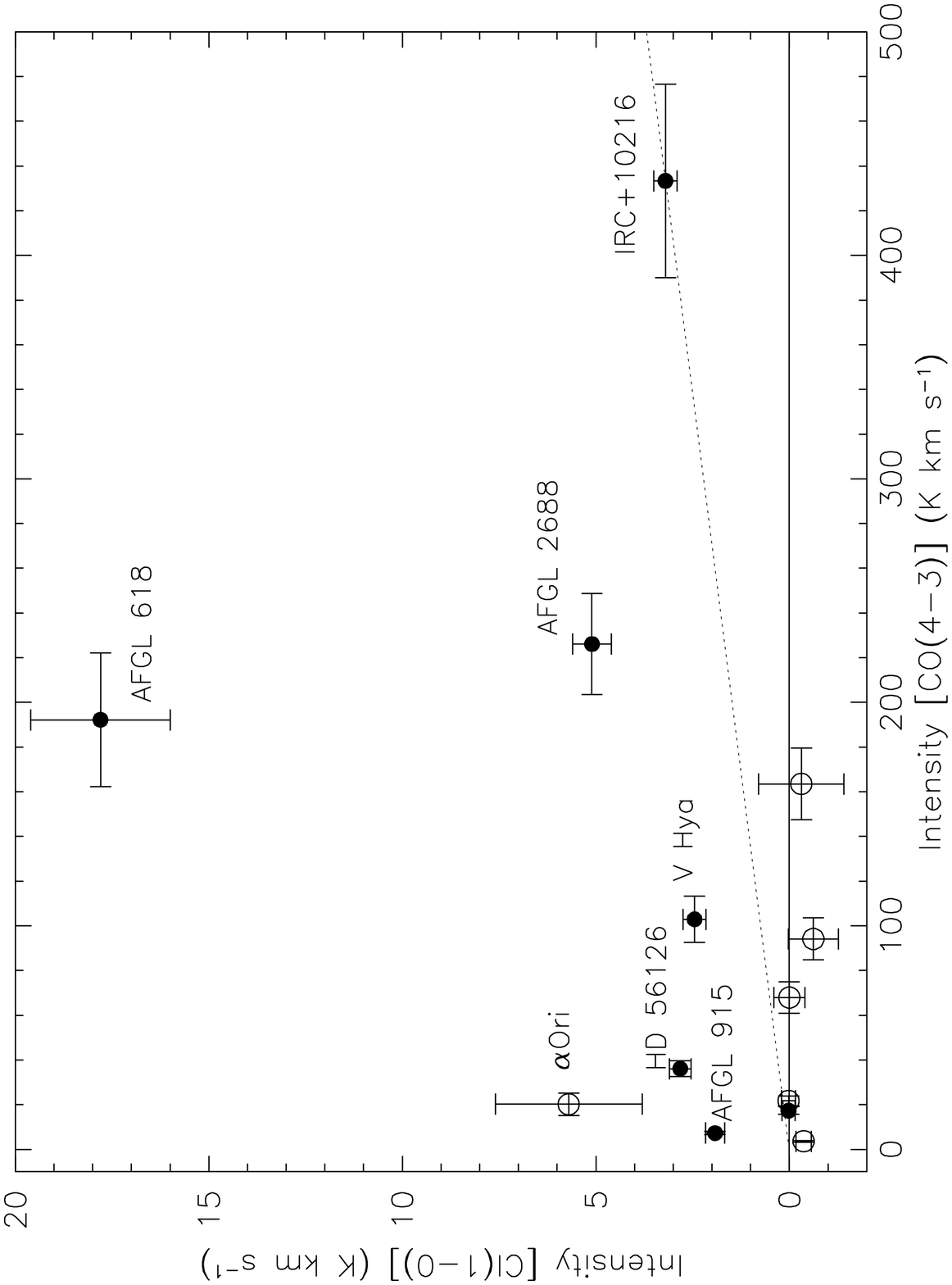}

\clearpage

\plotone{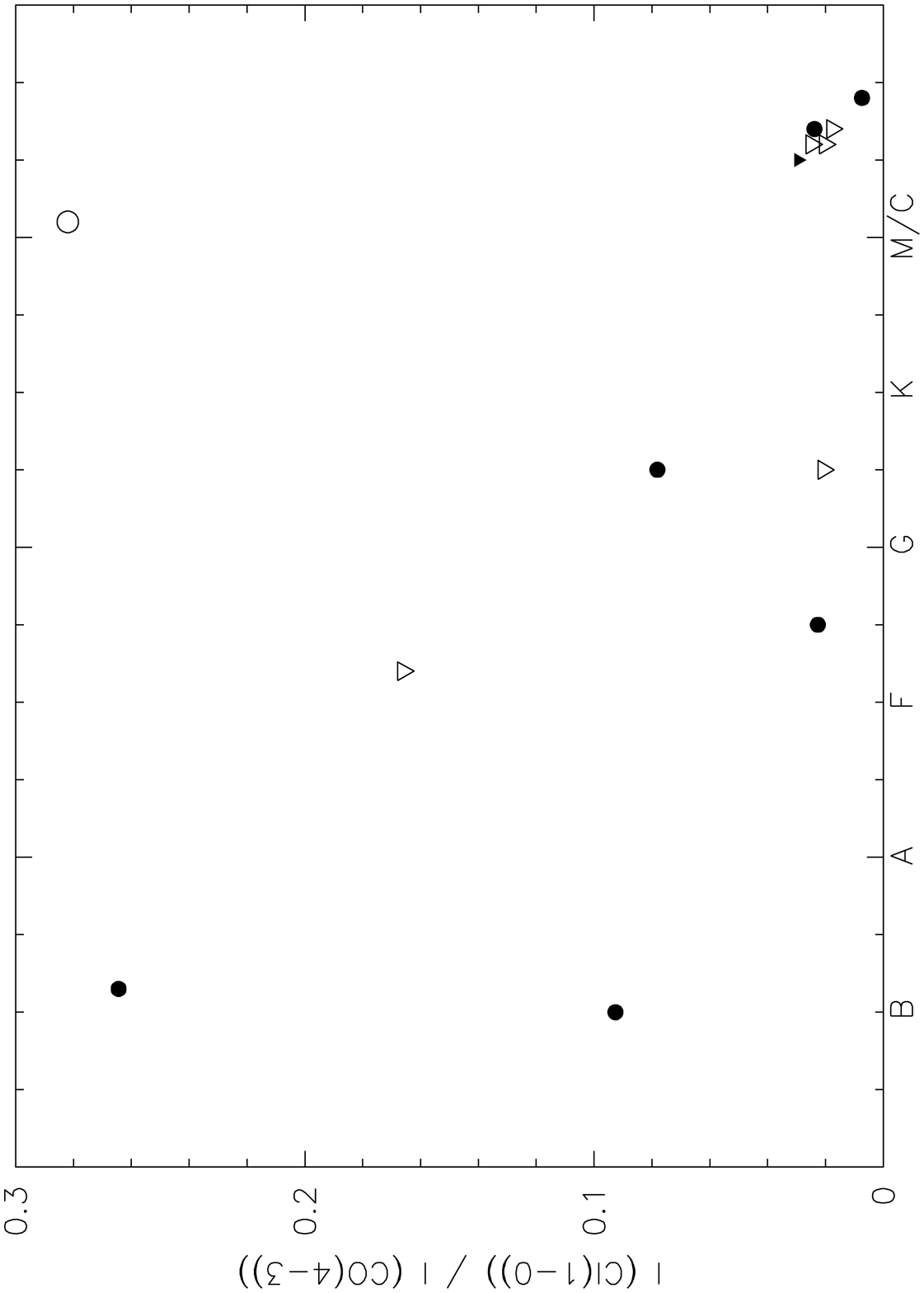}

\clearpage

\plotone{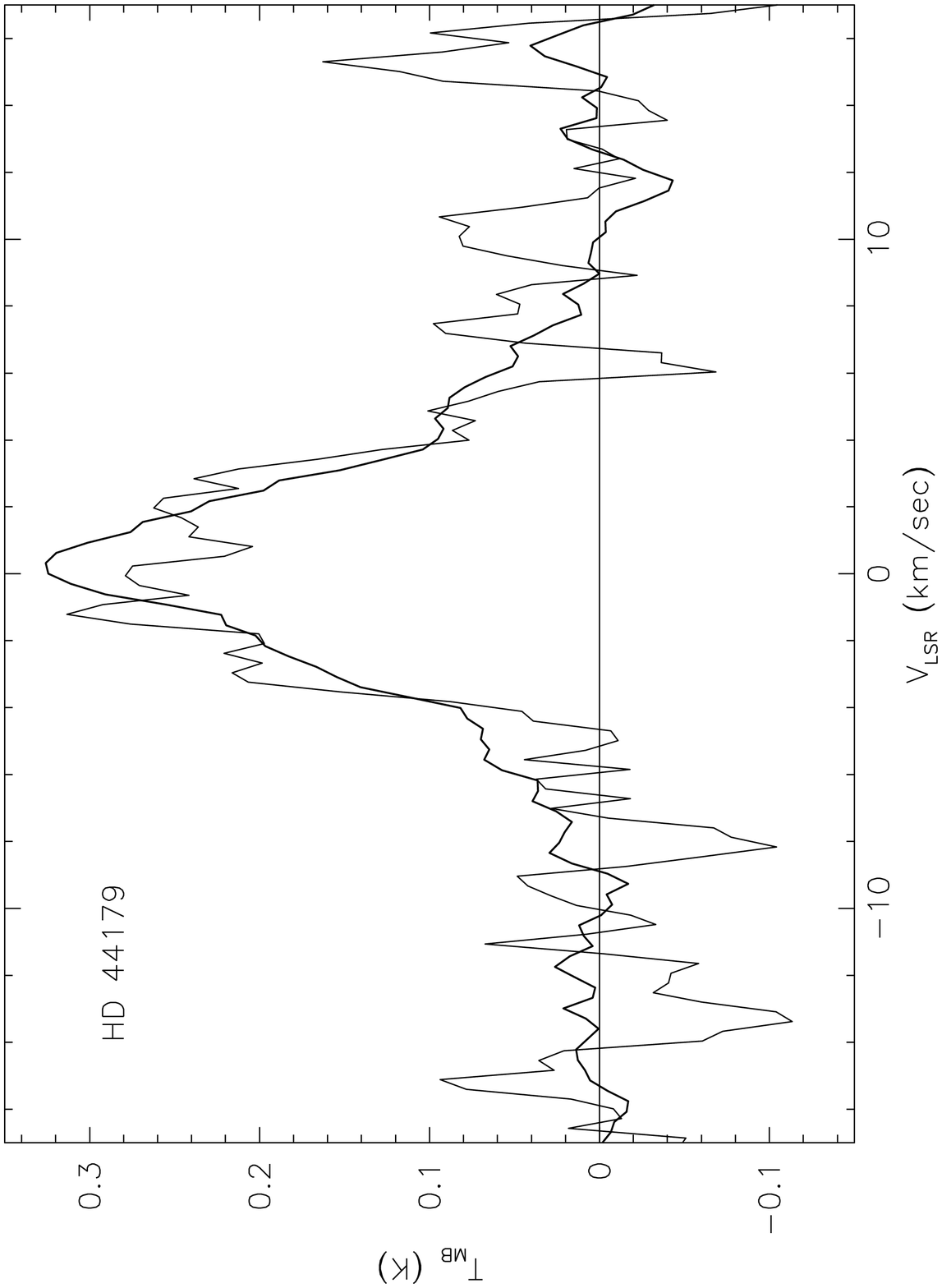}

\clearpage

\plotone{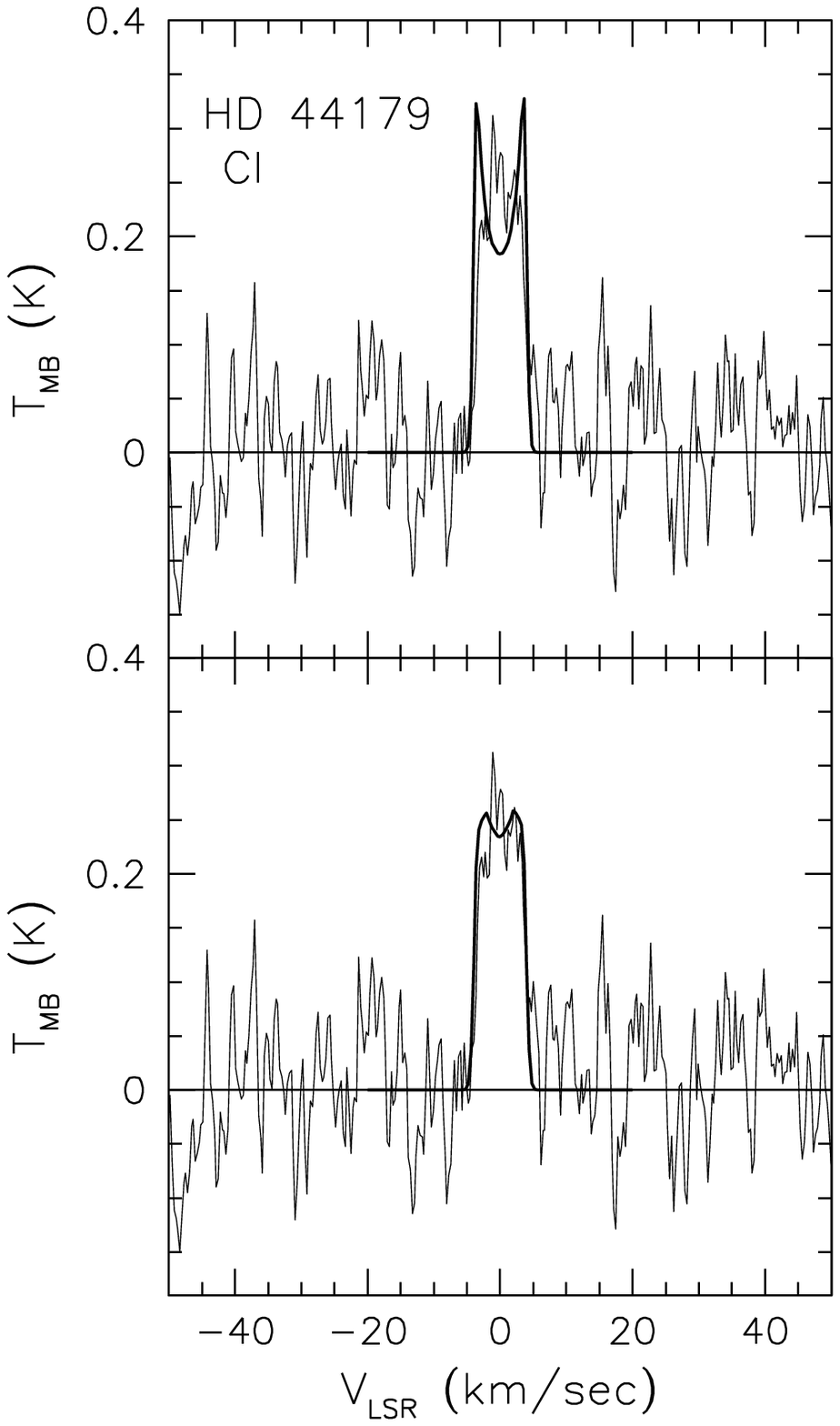}

\clearpage

\plotone{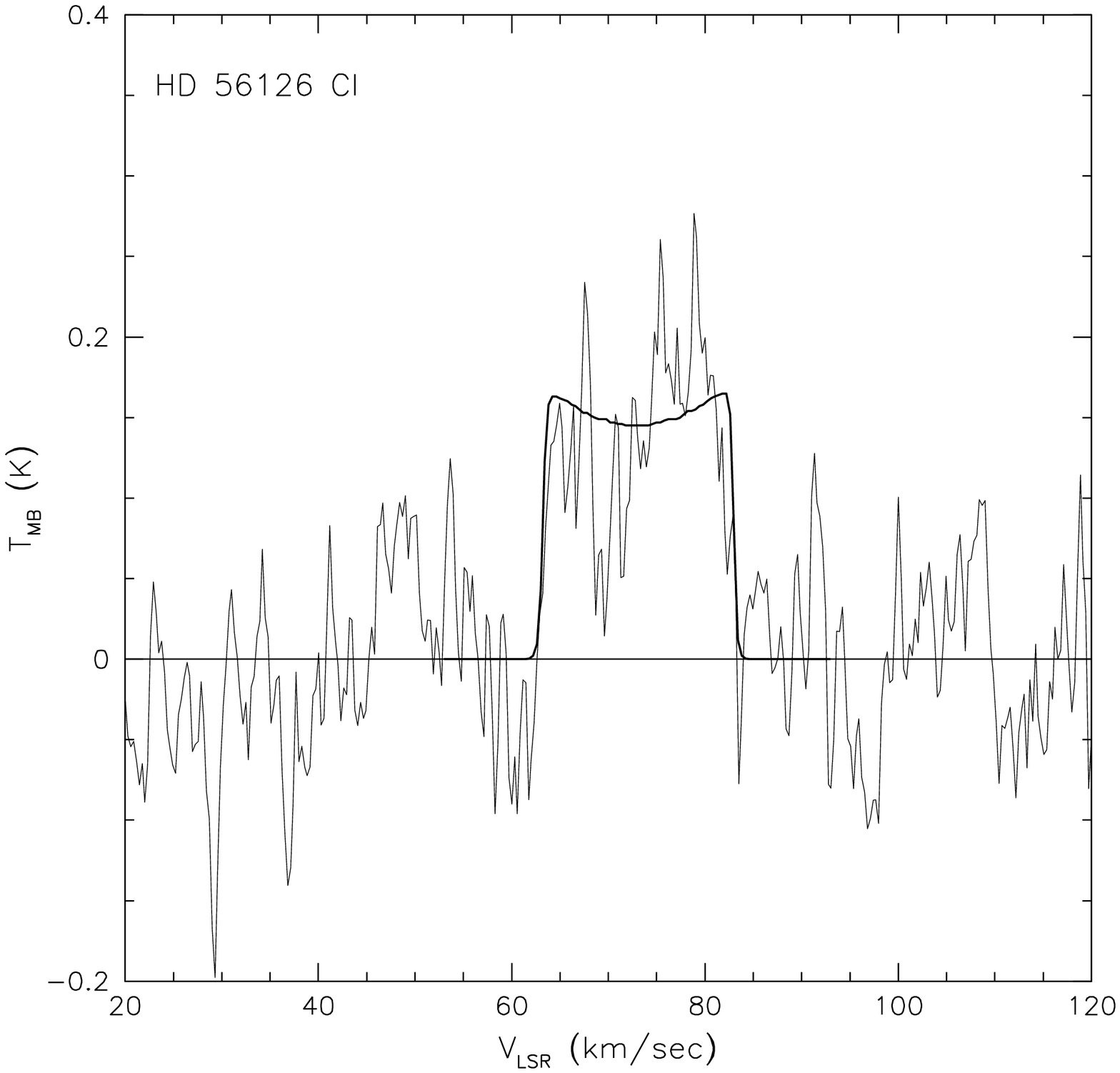}

\clearpage

\plotone{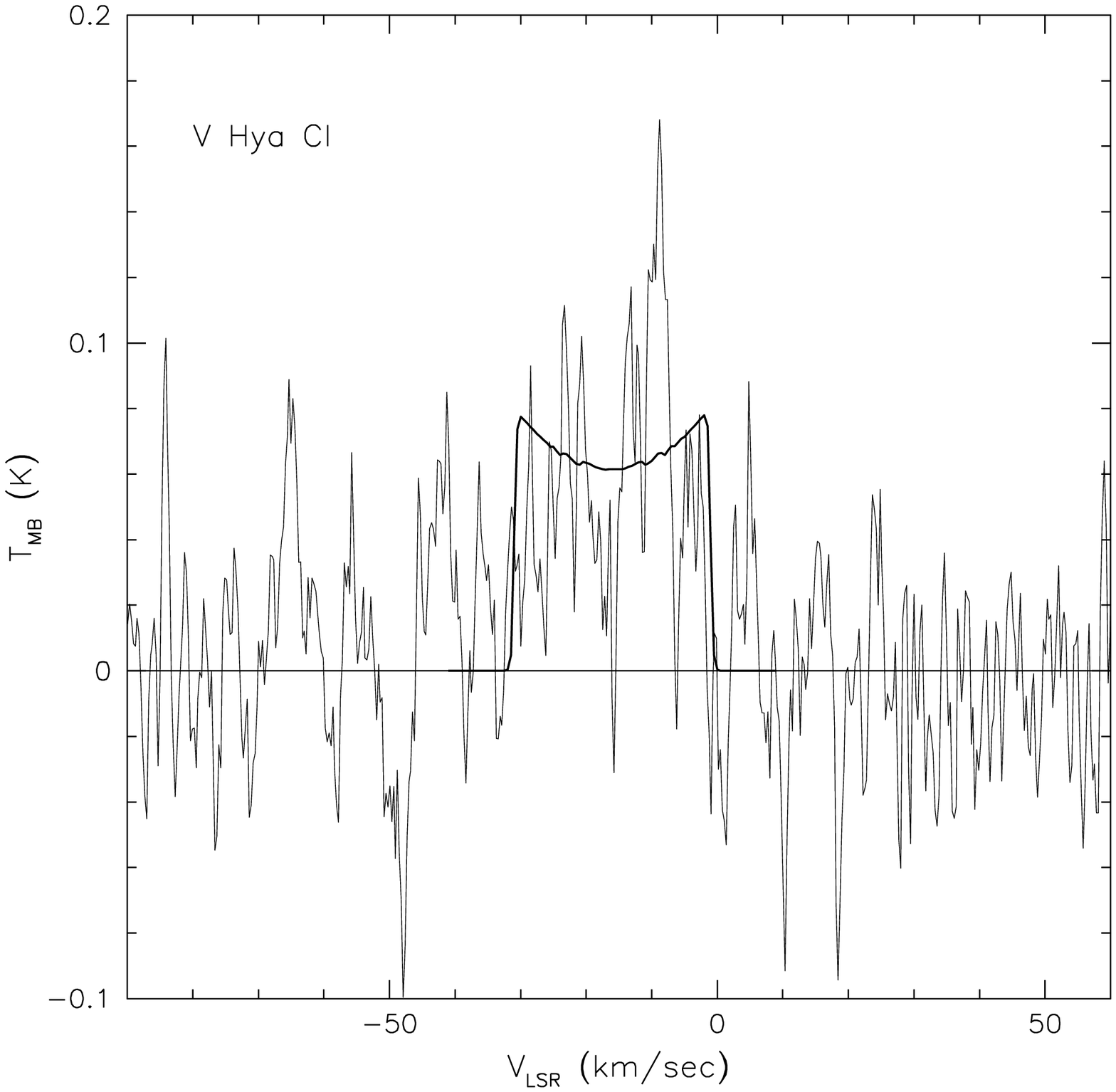}

\end{document}